\documentclass[12pt]{iopart}
\usepackage[latin9]{inputenc}
\usepackage{float}
\usepackage{graphicx}

\makeatletter

\providecommand{\tabularnewline}{\\}

\usepackage{iopams}
\usepackage{setstack}

\maketitle

\usepackage{color}

\makeatother

\begin{document}

\note[Microwave cavity tuned with liquid metal]{Microwave cavity tuned with liquid metal and its application to
Electron Paramagnetic Resonance}

\author{C.S.Gallo$^{1}$, E.Berto$^{2}$, C.Braggio$^{2,3}$, F.Calaon$^{3}$,
G.Carugno$^{2,3}$, N.Crescini$^{1,3}$, A.Ortolan$^{1}$, G.Ruoso$^{1}$,
M.Tessaro$^{3}$}

\address{$^{1}$ INFN, Laboratori Nazionali di Legnaro, Viale dell\textquoteright Università
2, 35020 Legnaro, Padova, Italy}

\address{$^{2}$ Dipartimento di Fisica e Astronomia, Via Marzolo 8, 35131
Padova, Italy}

\address{$^{3}$ INFN, Sezione di Padova, Via Marzolo 8, 35131 Padova, Italy}

\ead{carmelo.gallo@lnl.infn.it}

\submitto{Measurement Science and Technology}

\noindent{\it Keywords\/}: {microwave cavity, tunable resonance, liquid metal, Electron Paramagnetic
Resonance, GaInSn}
\begin{abstract}
This note presents a method to tune the resonant frequency $f_{0}$
of a rectangular microwave cavity. This is achieved using a liquid
metal, GaInSn, to decrease the volume of the cavity. It is possible
to shift $f_{0}$ by filling the cavity with this alloy, in order
to reduce the relative distance between the internal walls. The resulting
modes have resonant frequencies in the range $7\div8\,$GHz.
The capability of the system of producing an Electron Paramagnetic
Resonance (EPR) measurement has been tested by placing a 1 mm diameter
Yttrium Iron Garnet (YIG) sphere inside the cavity, and producing
a strong coupling between the cavity resonance and Kittel mode.
This work shows the possibility to tune a resonant system in the GHz
range, which can be useful for several applications.
\end{abstract}
\maketitle
Resonant cavities are typically completely enclosed by conducting
walls that can contain oscillating electromagnetic fields. The resonant
frequency $f_{0}$ of a mode in a rectangular cavity depends on the
distances between the internal surfaces of the walls. Imposing the
boundary conditions on the electromagnetic field trapped inside the
cavity, it is possible to obtain an analytic expression of the resonant
frequency $f_{0}^{m,n,l}$ for the $m,n,l$ mode. If we call $a,b,d$
the dimensions of a cavity filled with vacuum, this calculation yields
\begin{equation}
f_{0}^{m,n,l}=\frac{c^{2}}{2}\sqrt{\Big(\frac{m}{a}\Big)^{2}+\Big(\frac{n}{b}\Big)^{2}+\Big(\frac{l}{d}\Big)^{2}},\label{eq:1}
\end{equation}
where $c$ is the speed of light in vacuum \cite{key-1}. Eq.(\ref{eq:1})
states that a cavity resonates at frequencies which are determined
by the dimensions of the resonant cavity: as the cavity dimensions
increase, the resonant frequencies decrease, and vice versa. Thus
a reduction of one of this distances necessarily results in increased
resonant frequencies of the modes, allowing a tuning of the cavity
within certain ranges.

\begin{figure}[H]
\centering \includegraphics[width=0.9\textwidth,height=7cm]{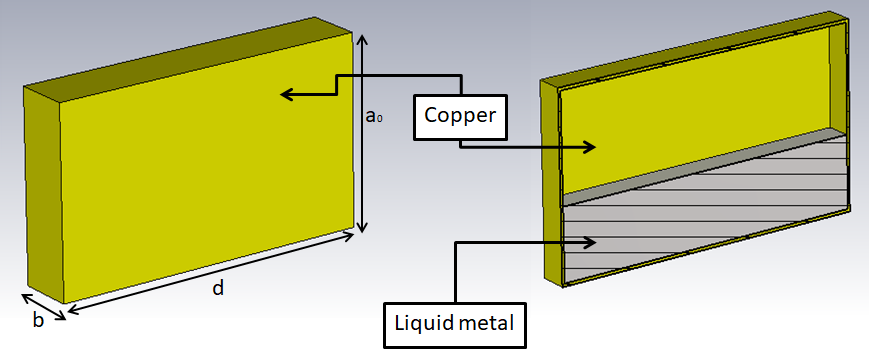}
\caption{\textit{Left}: the cavity. \textit{Right}: half view of the cavity
with liquid metal.}
\label{cavity:mode1} 
\end{figure}

To shift the resonance frequency of a chosen mode we change one single
dimension by filling the cavity with a liquid metal. It is to be noticed
that this is not the only way to shift the resonance frequency of
a mode, for example it is possible to insert dielectric materials
into the cavity \cite{key-2,key-3,key-4,key-5,key-6,key-7,key-8}.
The copper cavity used in this work has dimensions $(a = 30\,\mathrm{mm})\times(b
= 10\,\mathrm{mm})\times(d = 60\,\mathrm{mm})$; we aim to shift the resonance frequency of
a Transverse Magnetic mode (TM102), whose resonance frequency $f_{0}$
is 7.093 GHz. The cavity partially filled with liquid metal is shown
in Fig.(\ref{cavity:mode1}).

\begin{figure}[H]
\centering \includegraphics[width=0.9\textwidth,height=8cm]{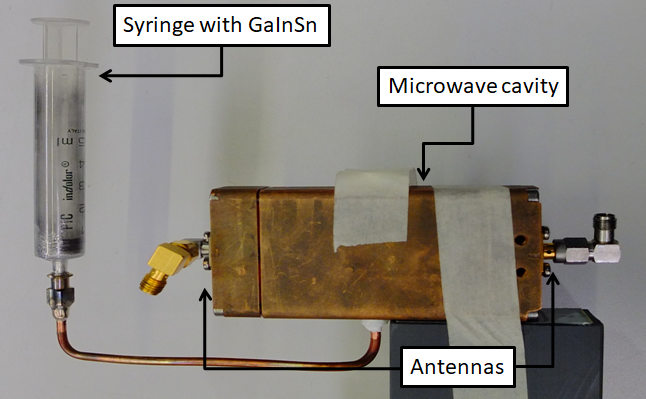}
\caption{Picture of the microwave cavity connected to the GaInSn syringe by
means of a copper tube soldered to the bottom of the cavity itself.}
\label{cavity_ph} 
\end{figure}

Let us call $a_{0}$, $f_{0}$ and $a_{i}$, $f_{i}$ the
coordinate in the vertical direction and resonant frequencies of the empty and filled cavity,
respectively. We tune the resonant frequency of the selected mode
injecting different volumes of a liquid metal. We have selected GaInSn
(liquid metal at room temperature), which is an eutectic mixture of
the metals gallium, indium and tin. We control the volume of the injected
metal using a syringe connected to the cavity by a copper tube as
shown in Fig.(\ref{cavity_ph}). The dimensions of copper tube are
not important, in our experiment we used a copper tube 130 mm long
and with 1.78 mm of diameter.

\begin{figure}[H]
\centering\includegraphics[scale=0.6]{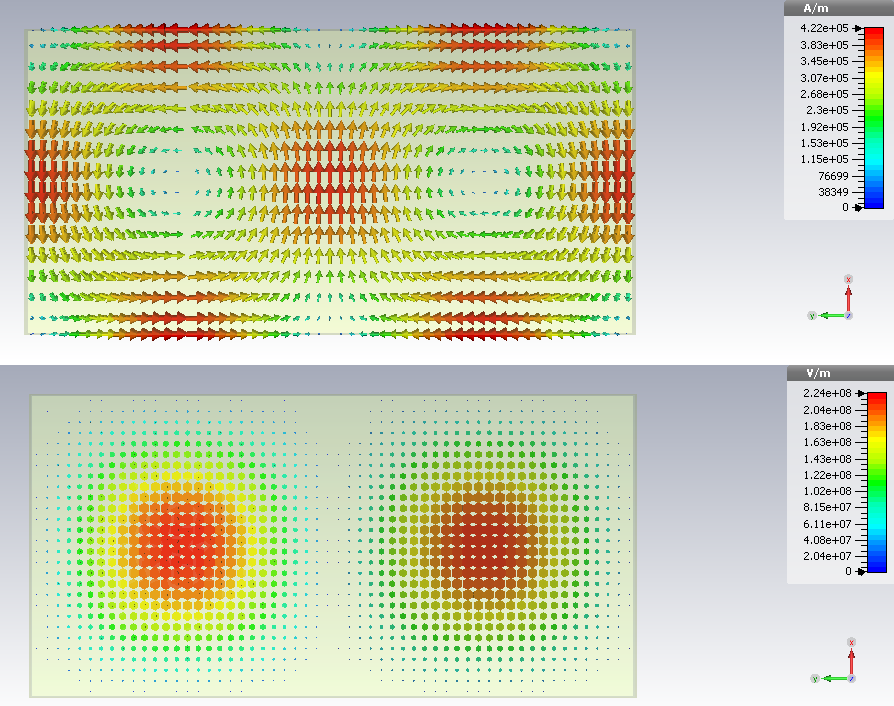} \caption{Simulation of the TM102 cavity mode fields: the top of the figure
represents the magnetic field, while the bottom part is the electric
field.}
\label{cavity:mode2} 
\end{figure}

\begin{figure}[H]
\centering\includegraphics[scale=0.7]{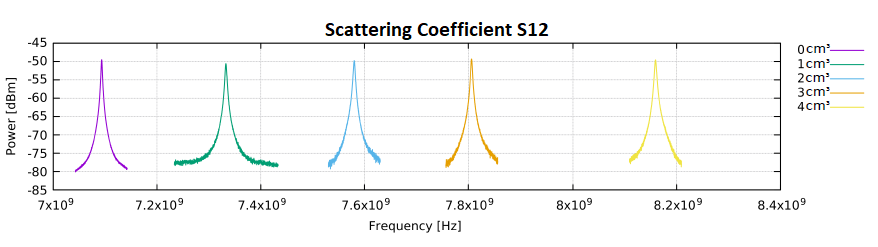} \caption{Module of Transmission Coefficient S12 of TM102 mode at different
volume of liquid metal.}
\label{S21} 
\end{figure}

The measures were made at room temperature. The measures consist of
$i=1,2,3,4,5$ steps. For every $i^{\mathrm{th}}$ step the
volume of liquid metal inside the cavity is increased of 1 cm$^{3}$.
In each step, we measured the scattering parameters of the
system using a Network Analyzer, which is connected to the microwave
cavity by two loop antennas (see Fig.(\ref{cavity_ph})). For each
variation of the cavity volume, we simulated the system by CST
Microwave Studio to find the resonant frequencies, and the electromagnetic field of the TM102 cavity mode. In Fig.(\ref{cavity:mode2})
we report an example of the fields in the cavity without liquid metal.
While in Fig.(\ref{S21}) we show the measurements of the module of
the transmission coefficient (S12) at different volume of liquid metal,
taken with the Network Analyzer.

Now we were able to compare the frequencies measured with the simulated
ones. The results are reported in Tab.(\ref{tab_res}).

\begin{table}[H]
\caption{\label{tab_res}Results of the different measurements and simulations.
The columns represents the cm$^{3}$ of metal injected into the cavity,
the residual volume of the partially filled cavity, the simulated
($f_{i}^{\mathrm{th}}$) and the measured ($f_{i}$) resonance frequencies,
and the corresponding measured loaded $Q$ factor.}
\begin{indented} 

\centering \item%
\begin{tabular}{@{}lccccl}
\br {[}cm$^{3}${]}  & $a_{i}\times b\times d$ {[}mm$^{3}${]}  & $f_{i}^{\mathrm{th}}$ {[}GHz{]}  & $f_{i}$ {[}GHz{]}  & $Q_{i}^{(L)}$  & \tabularnewline
\mr 0  & 30.0$\times$10$\times$60  & 7.066  & 7.093  & 2728  & \tabularnewline
1  & 28.3$\times$10$\times$60  & 7.282  & 7.333  & 2156  & \tabularnewline
2  & 26.7$\times$10$\times$60  & 7.516  & 7.580  & 2106  & \tabularnewline
3  & 25.0$\times$10$\times$60  & 7.805  & 7.806  & 2296  & \tabularnewline
4  & 23.3$\times$10$\times$60  & 8.146  & 8.159  & 2039  & \tabularnewline
\br  &  &  &  &  & \tabularnewline
\end{tabular}\end{indented} 
\end{table}
 As expected, the resonant frequencies change in good agreement with
the simulations and the loaded quality factors $Q_{i}^{(L)}$ of the
modes does not differ drastically from those of the empty cavity.
We have thus shown that it is possible to tune the resonance frequencies
of a cavity filled with a liquid metal.

Now we demonstrate that the system can be used for a physical application
like an Electron Paramagnetic Resonance (EPR) experiment. In this way we tested
the coupling of the cavity with a ferromagnetic resonance of a magnetic material
placed inside. The cavity is equipped with a magnetized sample of
volume $V_{S}$ and then placed inside an electromagnet that generates
a static magnetic field $B_{0}$. In a ferromagnetic material, electron
spins tend to align parallel to the external magnetizing field $B_{0}$ 
and the Larmor frequency of the ferromagnetic resonance reads
$f_{L}=\frac{\gamma B_{0}}{2\pi}$, where $\gamma$ is the electron gyromagnetic ratio. 
If the total number of spin of
the sample is sufficient, when the Larmor frequency and the cavity
resonance coincide, the resonant mode of the system splits in two
separate modes (hybridization). The mode separation is given by the
total coupling strength $g=g_{0}\sqrt{n_{S}V_{S}}$, where $n_{S}$
is the spin density of the material, $V_{S}$ is the volume of the
material, $g_{0}=\gamma\sqrt{\frac{\mu_{0}\hslash\omega_{L}}{V_{C}}}$
is the single spin coupling, $V_{C}$ is the volume of the cavity
and $\omega_{L}=2\pi f_{L}$ \cite{key-9}. To measure the hybridization,
with the loop antennas, in addition to the field $B_{0}$, an RF field
$B_{1}$ is applied to the sample in a direction orthogonal to $B_{0}$.
We placed a 1\,mm diameter Yttrium Iron Garnet (YIG) sphere in the
center of the cavity, where the RF magnetic field $B_{1}$ is maximum
(see Fig.(\ref{cavity:mode2})). This material has very high spin
density $n_{S}=2\cdot10^{28}$ m$^{-3}$. The volume of the material
is $V_{S}=\frac{4}{3}\pi r^{3}=0.52\cdot10^{-9}$ m$^{-3}$, so the
total expected coupling strength is $g=3.24\cdot10^{9}\,g_{0}$.

\begin{figure}[H]
\centering \includegraphics[scale=0.7]{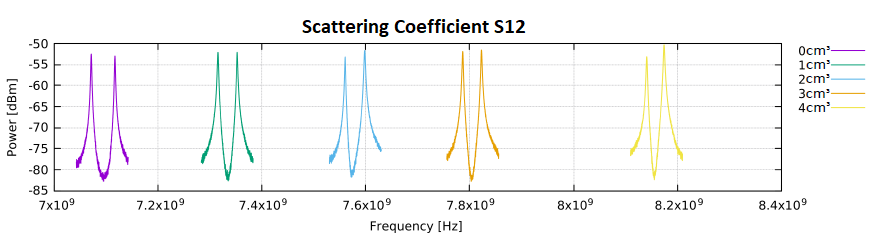}
\caption{Module of Transmission Coefficient S12 of hybrid resonances at different
volume of liquid metal.}
\label{all_res} 
\end{figure}

We also performed 5 measures, labelled with $j$, for the system cavity plus YIG. For
every $j=1,2,3,4,5$ step the volume of liquid metal inside the
cavity has been increased of 1 cm$^{3}$. We measured the scattering parameters of the system using the Network
Analyzer. The results of measurements are reported in Tab.(\ref{tab_resh}),
and in Fig.(\ref{all_res}) we show the measurements of the module
of the transmission coefficient (S12) of hybrid system at different
volume of liquid metal.

\begin{table}[H]
\caption{\label{tab_resh}Results of the different measurements in the strong
coupling regime. The columns represents the cm$^{3}$ of metal injected
into the cavity, the applied magnetic field, the lower ($f_{i}^{<}$)
and higher ($f_{i}^{>}$) hybrid resonances, and the respective Q
factors.}
\begin{indented} 

\centering \item%
\begin{tabular}{@{}llllll}
\br {[}cm$^{3}${]}  & $B_{z}$ {[}T{]}  & $f_{i}^{<}$ {[}GHz{]}  & $Q_{i}^{(L)<}$  & $f_{i}^{>}$ {[}GHz{]}  & $Q_{i}^{(L)>}$ \tabularnewline
\mr 0  & 0.253  & 7.071  & 4159  & 7.117  & 4448\tabularnewline
1  & 0.262  & 7.315  & 3325  & 7.353  & 3501\tabularnewline
2  & 0.270  & 7.561  & 3979  & 7.599  & 2923\tabularnewline
3  & 0.279  & 7.778  & 3709  & 7.824  & 3556\tabularnewline
4  & 0.291  & 8.141  & 3540  & 8.174  & 3027\tabularnewline
\br  &  &  &  &  & \tabularnewline
\end{tabular}\end{indented} 
\end{table}

 If we consider, for example, the cavity without liquid metal, the
single spin coupling is $g_{0}=0.016$ Hz, so the total coupling strength
is $g=5.18\cdot10^{7}$ Hz. The total coupling strength measured
is $g=4.60\cdot10^{7}$ Hz, which are comparable within about 10\%.
The system works as expected, and for all the different levels of
liquid metal we were able to obtain hybridization. This demonstrates
the capability of our tunable resonant system of performing EPR measurement.

In conclusion, in this note we introduce a new method to tune the
frequencies of the modes of cavities using liquid metal, and how it
can be exploited in an EPR application. This process has
been verified in the $7\div8$ GHz frequency range, however we can sweep
over different frequency ranges by changing the geometry of the empty cavity.
\ack{}{}

We wish to thank doctor Antonio Barbon for helpful discussions concerning
the experiment.

\newpage{}


\section*{References}
\begin{thebibliography}{1}
\bibitem[1]{key-1}Pozar D 2004 Microwave Engineering (Wiley) ISBN
9780471448785 

\bibitem[2]{key-2}Liu X, Katehi L P B, Chappell W J and Peroulis
D 2010 Journal of Microelectromechanical Systems 19 774-784 ISSN 1057-7157

\bibitem[3]{key-3}Stefanini R, Chatras M, Pothier A, Orlianges J
C and Blondy P 2009 High q tunable cavity using dielectric less rf-mems
varactors 2009 European Microwave Integrated Circuits Conference (EuMIC)
pp 391-394

\bibitem[4]{key-4}Perigaud A, Pacaud D, Delhote N, Tantot O, Bila
S, Verdeyme S and Estagerie L 2016 Frequency-tunable microwave-frequency
wave filter with a dielectric resonator including at least one element
that rotates uS Patent 9,343,791

\bibitem[5]{key-5}E K 1969 Tunable microwave cavity using a piezoelectric
device uS Patent 3,471,811

\bibitem[6]{key-6}C Carvalho N, Fan Y and Tobar M 2016 Review of
Scientic Instruments 87 094702

\bibitem[7]{key-7}Sakaguchi J, Gilg H, Hayano R, Ishikawa T, Suzuki
K, Widmann E, Yamaguchi H, Caspers F, Eades J, Hori M, Barna D, Horvth
D, Juhsz B, Torii H and Yamazaki T 2004 Nuclear Instruments and Methods
in Physics Research Section A: Accelerators, Spectrometers, Detectors
and Associated Equipment 533 598 \textendash{} 611 ISSN 0168-9002
URL http://www.sciencedirect.com/science/article/pii/S0168900204014639

\bibitem[8]{key-8}Carter P S 1961 IRE Transactions on Microwave Theory
and Techniques 9 252\textendash 260 ISSN 0097- 2002

\bibitem[9]{key-9}Tabuchi Y, Ishino S, Ishikawa T, Yamazaki R, Usami
K and Nakamura Y 2014 Phys. Rev. Lett. 113(8) 083603
\end{thebibliography}
\end{document}